\documentclass{Rinton-P9x6}

\begin{document}

\title{CYCLOTOMIC QUANTUM CLOCK}

\author{Haret C. Rosu}

\address{Applied Mathematics, IPICyT, Apdo. Postal 3-74
Tangamanga, San Luis Potos\'{\i}, M\'exico
\\E-mail: hcr@ipicyt.edu.mx}

\author{M. Planat}

\address{Laboratoire de Physique et M\'etrologie des Oscillateurs du CNRS, 25044 Besan\c con Cedex, France\\
E-mail: planat@lpmo.edu}  

\address{Dated: August 2003; File: RP.tex}
\address{Proc. ICSSUR-8, Puebla, Mexico, 9-13 June 2003\\
Eds. H. Moya-Cessa, R. J\'auregui, S. Hacyan, O. Casta\~nos\\ Rinton Press, ISBN 1-58949-040-1, pp. 366-372 (Nov. 2003)}


\maketitle

\abstracts{In the wake of our recent work on cyclotomic effects in quantum phase-locking [M. Planat \& H. C. Rosu, Phys. Lett. A 315, 1 (2003)],
we briefly discuss here a cyclotomic extension of the Salecker \& Wigner quantum clock. We also hint on a possible cyclotomic structure of time
at the Planck scales.}


\section{Introduction}
Time can be defined in many different ways depending on context and discipline.\cite{zeman} In quantum physics its operatorial features that are required by the quantum formalism are a matter of ongoing debate.


Indeed, time appears to be merely a parameter ($c$-number) but of course
dynamical variables of specific quantum systems (clocks) may be used as \underline{substitutes of time}. In 1933, Pauli gave a famous 
argument on the impossibility of introducing a {\em universal} time operator $T_u$ obeying the canonical commutation relation $[T_u,H]=i\hbar$ for \underline{any} Hamiltonian. According to Pauli,\cite{P33} this is forbidden because {\em some} systems have discrete energy eigenvalues. This forces one to introduce a discrete time 
operator which is not universal because there are many
types of discrete spectra. Thus, as recently stated by Hilgevoord,\cite{hilge} a quantum clock system may display a continuous time variable, or a discrete time variable, or 
no time variable at all. 


\noindent
In this work, we introduce a cyclotomic approach for the quantum clock model of Salecker and Wigner (SW).\cite{sw58} We also comment on a possible 
cyclotomic structure of the Planck time.  


\section{The Salecker-Wigner (SW) Clock}

\medskip

We will follow Peres' discussion  of SW clock properties,\cite{peres} due to his introduction of the pointer states (see below) and a time 
projector operator for this case. By quantum clock one basically means a microscopic device able to measure time with an accuracy $\tau$ over a time lag $T=n\tau$. 
This minimal requirement has been imposed by
Salecker and Wigner in 1958 in the context of quantum limitations of the measurements of space-time distances that led to 
the `smallest clock' Wigner constraints.\cite{sw58} 
During the interval $T$, the quantum mechanical state of the clock goes through $N$ orthogonal vectors, equivalent to asserting that the wave function
of such a clock could be written as a superposition of $N$ stationary states $u _k$, $k=1,...N$

\begin{equation}\label{hosc}
\phi(t)=\sum _{1}^{N} a_ku _k e^{-i\omega _k t}~. 
\end{equation}
The energy values of the stationary states were denoted by $\hbar \omega _k$.
The wave function $\phi$ goes through the $N$ orthogonal states at times $\tau$, $2\tau$, ..., $N\tau$ if $a_k=N^{-1/2}$ and $\omega _k=\omega _0+2\pi k/N\tau=\omega _0+2\pi k/T$, see Eq.~(11) below. In general an odd number, $N=2j+1$, of clock states of the azimuthal type
\begin{equation}\label{st1}
u_n(\theta)=\frac{1}{\sqrt{2\pi}}e^{in\theta}~, \quad n=0,...,N-1~,\quad \theta \in [0,2\pi)~,
\end{equation}
are employed because of their simple Hamiltonian that implies an ordered time evolution. 

\medskip

\subsection{SWP Pointer States $v_k(\theta)$}

\medskip

Peres introduced another orthogonal basis of clock states based on the azimuthal states 

\begin{equation}\label{st2}
v_k(\theta)=N^{-\frac{1}{2}}\sum^{N-1} e^{-2\pi ikn/N}u_n(\theta)=
(2\pi N)^{-1/2}\frac{\sin\Big[\frac{N}{2}\left(\theta -\frac{2\pi k}{N}\right)\Big]}{\sin \Big[\frac{1}{2}\left(\theta -\frac{2\pi k}{N}\right)\Big]}
\end{equation}
for $k=0,..., N-1$. For large $N$, these functions have a sharp peak at
\begin{equation}
\theta _{\rm peak}=\frac{2\pi k}{N}
\end{equation}
and can be visualized as ``pointing to the $k$th hour" with an angle uncertainty $\pm \pi/N$. This is why we call these states pointer states. 

We can then define projection operators 
\begin{equation}
P_kv_m=\delta _{km}v_m~,
\end{equation}
and a ``clock time" operator
\begin{equation}
T_{c}=\tau \sum^{N-1} kP_k~,
\end{equation}
where $\tau$ is the time resolution of the quantum clock. 

The eigenvectors of $T_c$ are the functions $v_k$, and the corresponding eigenvalues are $t_k=k\tau$, with $k=0,..., N-1$. 
Therefore, measuring $T_c$ can at best yield a discrete approximation to the true time similar to an ordinary watch.
The initial state of the clock can be always assumed to be $v_0$.

The clock Hamiltonian will be written as
\begin{equation}
H_c=\omega J~,
\end{equation}
where $\omega=2\pi/N\tau$ and $J=-i\hbar \partial/\partial \theta$. Since
\begin{equation}
H_cu_m=m\hbar \omega u_m~,
\end{equation}
one gets
\begin{equation}
\exp(-iH_c t/\hbar)u_m=e^{-im\omega t}u_m=(2\pi)^{-1/2}e^{im(\theta -\omega t)}~.
\end{equation}

It follows that 
\begin{equation}
\exp(-iH_c\tau /\hbar)v_k=v_{k+1\,({\rm mod} N)}~,
\end{equation}
showing that the clock passes succesively through the states $v_0,v_1,v_2,...$ at time intervals $\tau$.

\medskip

\subsection{Commutation operator}

\medskip

\noindent
These discrete Hamiltonian and clock-time operators cannot satisfy $[T_c, H_c]=i\hbar$. A straightforward calculation gives
\begin{eqnarray} 
 \langle u_m|[T_c,H_c]|u_n \rangle \,= \, \langle 
v_m|[T_c,H_c]|v_n\rangle &=& \qquad \,\, 0 \qquad \qquad \qquad \quad \,\, \qquad (n=m) \nonumber\\
& = &i\hbar\frac{\frac{2\pi i}{N} (n-m)}{1-\exp[\frac{2\pi i}{N} (n-m)]} \quad (n\neq m)~.\nonumber
\end{eqnarray}
This rather complicated result is due to the discontinuity of the quantum clock time when going from the $2j$th to the zeroth hour eigenvalue.

\medskip

\subsection{Energetics of the SW clock}

\medskip
\noindent
The pointer eigenstates $v_k$ satisfy $\langle H_c\rangle=0$ and 
\begin{equation}
(\Delta H_c)^2=|H_cv_k|^2=\frac{\hbar ^2\omega ^2}{N}\sum^{\sim N} m^2\approx\hbar ^2 \omega ^2 N^2/3~.
\end{equation}

For large $N$, the energy uncertainty
\begin{equation}
\Delta H_c\approx N\hbar \omega /\sqrt{3}\approx (\pi/\sqrt{3})(\hbar /\tau)
\end{equation}
is almost as large as the maximum available energy $\approx N\hbar \omega$. This demonstrates that the clock $H_c$ is 
essentially a nonclassical object.







\bigskip

\section{Cyclotomic reformulation}

\medskip

\noindent 
A state space of finite dimension $N$ is also the basic framework of the Pegg \& Barnett (PB) quantum phase
formalism.\cite{pb} The PB phase
states are the equivalent of the $v_k$ pointer states. The difference is that the former are defined as superpositions of number states, whereas the latter 
are similar types of superpositions but of azimuthal states $u_n$. Thus, we can apply our recent cyclotomic procedure in quantum phase-locking to the quantum
clock problem. We write the SWP pointer states in the form
\begin{equation}
|v _p\rangle=N^{-1/2}\sum_{n=0}^{N-1}e_N^p(-n)|u_n\rangle~,
\label{eq1}
\end{equation}
where
\begin{equation}
e_N^p(-n)=\exp(-2i\pi\frac{p}{N}n)~ \label{eq1b}
\end{equation}
is the $n^{\rm th}$ power of the $p^{\rm th}$ 
root of the unity in the ordered set from $0$ to $N-1$; in group theoretical terms it is the character of
the group $G=\mathcal{Z}/N\mathcal{Z}$. Notice that $p$ is a fixed
number in the set (0...$N-1$). 
The states $|v _p\rangle$ can be considered as
quantum {\em  inverse} discrete Fourier transforms on the set of azimuthal
states. They are inverse Fouriers because of the minus sign in $e_N^p$. 
Once this noticed, one can proceed with the cyclotomic extension of the 
model in parallel with our recent work on quantum phase-locking.\cite{pla}  



%
%

%

\medskip

\subsection{Time operator with coprimality}

\medskip

\noindent We can define a cyclotomic time operator
\begin{equation}
T_{cyclot}=\tau\sum_{\stackrel{p=0}{(p,N)=1}}^{N-1}p
|v _p\rangle\langle v
_p|~,
\label{eqop}
\end{equation}
where $(p,N)=1$ means that the sum is taken over all $p$ coprime
with $N$.

\bigskip

\subsection{Cyclotomic canonical commutator}

\medskip

\noindent The calculation of the cyclotomic SW time-hamiltonian commutator leads to
the following expression

\begin{equation}
[T_{cyclot},
H_{c}]=\frac{\tau}{N}\sum_{\stackrel{p,n,l=0}{(p,N)=1}}^{N-1}p(l-n)
e_N^p(-n)e^{*p}_{N}(-l)|v_n\rangle\langle v_l|~.
\label{eqcom1}
\end{equation}

The matrix elements of the cyclotomic canonical commutator in the
clock states are given by
\begin{eqnarray}
\langle v_n|[T_{cyclot}, H_c]|v_n\rangle
&=&0~,\nonumber\\
\langle v_l|[T_{cyclot}, H_c]|v_n\rangle
&=&\frac{\tau}{N}\sum_{\stackrel{p=0}{(p,N)=1}}^{N-1}p(l-n)
\exp\left(\frac{2i\pi p}{N}(l-n)\right)~.
\end{eqnarray}

\bigskip

\subsection{Commutator for a general superposition of azimuthal states}

\medskip

\noindent Let us consider now a general superposition state of
azimuthal states
\begin{equation}
|f\rangle=\sum_{n=0}^{N-1}c_n|u_n\rangle~. \label{eqgs1}
\end{equation}
The expansion in SW pointer states will be
\begin{equation}
|f\rangle=\frac{1}{N}\sum_{n,p}^{N-1}c_n\exp(-2i\pi\frac{p}{N}n)|v
_p\rangle~. \label{eqgs2}
\end{equation}

The expectation value of the cyclotomic canonical commutator in the $|f\rangle$ states will be
\begin{equation}
\langle f|[T_{\rm cyclot},H_c]|f\rangle
=\frac{\tau}{N}\sum_{\stackrel{n,l}{(p,N)=1}}^{N-1}c_{n}^{*}c_l
a_{ln}~, \label{eqgs3}
\end{equation}
where
\begin{equation}
a_{ln}=\sum_{\stackrel{p}{(p,N)=1}}^{N-1}(l-n)p
\exp\left(\frac{2i\pi p}{N}(l-n)\right)~.\label{eqgs4}
\end{equation}

\bigskip

\section{Cyclotomic time at Planck scale ?}

In this section, we speculate on a possible cyclotomic structure of time at the Planck `quantum' scale $t_P\approx 5.4 \times 10^{-44}$ s.

There is a wealth of recent literature on observable consequences of Planckian fluctuations, i.e., quantum fluctuations 
of general relativity. Especially interesting is the recent debate on whether the usual Airy rings form in the case of 
distant compact radiation sources, such as active galactic galaxies and gamma ray bursts.\cite{astro}

We think that a detailed investigation of cosmic diffraction could reveal a fundamental cyclotomic structure. 
Cyclotomy of the Planckian time would lead to fingerprints in 
the phase of the high-energy electromagnetic pulses of astrophysical and/or cosmological origin. In other words, the temporal diffraction of such
pulses on the cyclotomic structure of the Planck time could  produce revivals like in ordinary quantum mechanics. 
In his 1996 RMP colloquium on string duality,\cite{pol96} Polchinski tackled higher dimensions as waveguides and discussed a Klein-Gordon equation in five dimensions as
an equivalent of a corresponding Helmholtz equation. Taking more seriously this optical analogy, one can elaborate on both spatial and temporal self-imaging
(Talbot) effects within this Planckian waveguides of cross section  $10^{-66}$ cm$^2$ similarly to Rohwedder's recent study on self-imaging in atom waveguides.\cite{roh01}
What might be the origin of a fundamental cyclotomic structure ? In the case of dynamical events of measurement type the origin is known to be phase-locking.
Since quantum gravity is essentially a nonlinear theory one should imagine a sufficiently general framework for a nonlinear phase-locking phenomenon such as autoresonance.\cite{f98} A resonant driving of a quantum field relativistic Bose-Einstein condensate could be a reasonable option.  

Before concluding, it is worth mentioning that cyclotomic features at the Planckian scale revealed by a Talbot type effect are not the only way to introduce number-theoretic ideas at this ultimate scale. A sort of wave-particle complementarity (i.e., between the Helmholtz equation and Fresnel integral) for the Talbot effect has been related
to the so-called reciprocity property of Gauss sums in number theory by Matsutani and $\hat{\rm O}$nishi.\cite{mo} Interestingly, coprimality conditions are also important
in Gauss sum reciprocity.


\section{Conclusion}

The concept of cyclic properties of time in quantum mechanics could, in principle, foster interesting ideas.
We introduced a cyclotomic modification of the common definition of the time operator as a projector in the space of the pointer states of the Salecker-Wigner quantum clock. 
Finally, we commented on a possible cyclotomic structure of time at the Planck scales.  


\footnotesize{}

\bigskip
\bigskip

\begin{center}

\noindent
{\footnotesize \em In a cyclical concept of time every starting point will have to be an ending point. 

\noindent
The notion of cyclical time is common to religions like Buddhism and Hinduism.

\noindent
Indian mythology refers to cyclical time as `anaadi' or that without a beginning.

\noindent
Cyclotomy is the counterpart of `anaadi' in Mathematics.}  

\end{center}

\end{document}